\documentclass[aps, pra, superscriptaddress,twocolumn, longbibliography]{revtex4-2}
\usepackage{soul}
\usepackage{graphicx}% Include figure files
\usepackage{bm}% bold math
\usepackage[normalem]{ulem}
\usepackage{amsbsy,amsmath}

\usepackage{palatino}
\usepackage[utf8]{inputenc}
\usepackage{epsf}
\usepackage{amsmath,amssymb}
\usepackage{calc}
\usepackage{amsmath,amssymb, bm}
\usepackage{float}

\usepackage[usenames,dvipsnames]{xcolor}

\usepackage{graphicx}
\usepackage{hyperref}
\usepackage[usenames,dvipsnames]{xcolor}

\newcommand{\ben}{\begin{equation}}
\newcommand{\een}{\end{equation}}
\newcommand{\bea}{\begin{eqnarray}}
\newcommand{\eea}{\end{eqnarray}}
\def\bra#1{\langle#1\vert}
\def\ket#1{\vert#1\rangle}

\def\sss{\scriptscriptstyle\rm}

\def\1s{_{1,\sss S}}
\def\2s{_{2,\sss S}}

\def\s{_{\sss S}}
\def\xc{_{\sss XC}}

\def\Hxc{_{\sss HXC}}

\def\H{_{\sss H}}

\def\ext{_{\rm ext}}

\def\br{{\bf r}}

\begin{document}

\title{Post-pulse dipole instability in adiabatic TDDFT: fact or artifact?}

\author{Davood B. Dar}
\affiliation{Department of Physics, Rutgers University, Newark 07102, New Jersey USA}
\affiliation{Department of Physical and Environmental Sciences, University of Toronto, Canada}
\author{Dhyey Ray}
\affiliation{Department of Physics, Rutgers University, Newark 07102, New Jersey USA}
%\author{Phuong Mai Dinh ?}
%\affiliation{Toulouse..}
%\author{Paul-Gerhard Reinhard ?}
%\affiliation{Erlangen..}
%\author{Eric Suraud ?}
%\affiliation{Toulouse..}
%\author{Tchavdar N. Todorov ?}
%\affiliation{Belfast..}
\author{Neepa T. Maitra}
\affiliation{Department of Physics, Rutgers University, Newark 07102, New Jersey USA}
\email{neepa.maitra@rutgers.edu}

\date{\today}

\begin{abstract}
Recent real-time TDDFT calculations have reported an unexpected delayed growth of molecular dipole oscillations some time after an extreme-ultraviolet (XUV) pulse is applied. We show that  numerical and analytical arguments suggest that this instability is an artifact of an incorrect non-linearity introduced by the computational approach: Propagation with an adiabatic exchange-correlation approximation within the time-dependent Kohn-Sham equations of time-dependent density functional theory (TDDFT) tends to amplify initially small and pure sinusoidal oscillations in a system. On the other hand, when this same adiabatic approximation is used within the recent response-reformulated RR-TDDFT, the instability is absent. The absorbing boundary condition plays a crucial role consistent with our argument.    We demonstrate this explicitly on the N$_2$ molecule subject to an XUV pulse, with a range of adiabatic functionals. 
\end{abstract}

\maketitle
\section{Introduction}
Predictive computational modeling of molecular processes far from their equilibrium requires a reliable method to capture both correlated electronic motion as well as coupling to ions. At the shortest timescales the electronic dynamics is of prime importance, and significantly influences the subsequent motion of the heavier nuclei. Even neglecting the nuclear motion, solving the full time-dependent Schr\"odinger equation (TDSE) for the many-electron wavefunction is impossible beyond a few electrons, and one typically resorts to reduced descriptions and approximations of interaction terms. Time-dependent density functional theory (TDDFT) has emerged as the most practical approach for such problems, which in principle captures the exact time-evolving density of the physical system via a set of non-interacting Kohn-Sham (KS) orbitals evolving in a one-body potential~\cite{RG84,Carstenbook,M16}. A key component in this potential is the exchange-correlation potential, $v\xc[n; \Psi(0),\Phi(0)](\br, t)$, a functional of the one-body density $n(\br,t'<t)$ (including its past evolution), the interacting initial state $\Psi(0)$ and the choice of initial Kohn-Sham state $\Phi(0)$. This is usually approximated by an adiabatic approximation, within which the instantaneous density is input in a chosen ground-state approximation, $v\xc^{\rm A}[n; \Psi(0),\Phi(0)](\br,t) = v\xc^{\rm g.s.}[n(t)](\br)$, thus neglecting all the memory-dependence. When orbital-dependent ground-state functionals such as hybrid functionals are used, the corresponding potential depends on the instantaneous orbital, and so has some memory in a limited sense.
The resulting adiabatic time-dependent Kohn-Sham (TDKS) equations 
\ben
i\,\partial_t \phi_i(\mathbf r,t) = \left[-\tfrac{1}{2}\nabla^2 + v\s(\mathbf r,t)\right]\phi_i(\mathbf r,t),
\label{eq:tdks}
\een
with $v\s= v\ext + v\H[n] + v\xc^{\rm A}[n,\Psi(0),\Phi(0)]$ as the sum of the external, Hartree, and adiabatic exchange-correlation potential, 
have been applied to simulate a range of photo-excited and laser-driven  phenomena in molecules and materials at the atto- and femto-timescales~\cite{LGIDL20,S21,SZYYK21,BCV22,U25,M16,LM23}.

One such domain of application is the use of ultrashort XUV pulses to investigate detailed time-resolved  ultrafast dynamics in molecules and clusters, with the promise to reveal the role of correlations and mechanisms in a number of intriguing processes, from Auger decay to charge-migration~\cite{CHL17,YUG18}.
    A recent series of papers have uncovered an unusual but persistent phenomenon predicted by these simulations~\cite{RDDSV23,RDDVS23,HDDVRS23,HDDRS25,DRDHS26}: a resurgence of a dipole signal some femtoseconds after an ultrafast XUV pulse centered at an off-resonant frequency is turned off. This signal, which has been termed a ``dipole-instability",  appears after a period of quiet, almost zero, dipole, that begins to grow with an exponential envelope before reaching a maximum, decaying, and repeating.  Such an instability could be measured in principle by time-resolved photo-emission spectroscopy, with a delayed ionization peak corresponding to the resurgence of the dipole.  The functional approximation used in the calculations satisfies several exact conditions, including the zero-force theorem,  and is energy-conserving,  and the dynamics are robust with different choices of numerical parameters. The phenomenon persists across a range of molecules, field strengths, frequencies, and orientations and survives with nuclear motion~\cite{HDDVRS23,RDDVS23}. The authors have highlighted that the key role of the pulse is in creating a population inversion in the system, and demonstrated that the instability also arises from instead creating an instantaneous hole in a deep-lying state; the dependence on the degree of depletion of this state has also been studied~\cite{HDDVRS23,RDDVS23,DRDHS26}. The population inversion (even partial) has been central in the analyses, where coupling of the upper state to many modes (continuum) is argued to result in a spontaneous symmetry-breaking, backed up by the presence of maxima in phase-space potential energy surfaces near the hole state~\cite{DRDHS26}.
While significant insight has been gained, particularly in the drilled hole scenario, the question has remained whether the observed instability is in fact real, or whether it is an artifact of the approximations inherent in the adiabatic TDKS equations. 

Here, we answer this question by considering two different formulations of TDDFT for real-time non-equilibrium dynamics: one is the traditional TDKS approach of Eq.~\ref{eq:tdks}, and the other is the recent response-reformulated TDDFT (RR-TDDFT)~\cite{DBM24} in which all exchange-correlation quantities needed can be obtained from ground-state or response calculations. While both are exact and would give identical results in the formal case where the exact functionals are used, RR-TDDFT has been demonstrated to be more accurate when the adiabatic approximation is used.  Using finite-basis set calculations, we reproduce the instability in the case where it develops after an XUV pulse is applied to the molecule in the TDKS simulation, and we find it is absent in the RR-TDDFT simulation.

We begin in Section~\ref{sec:RR-TDDFT} with a recall of the RR-TDDFT approach. In Sec~\ref{N2_dynamics}, we study the dipole dynamics of N$_2$, motivated by the case studied in Ref.~\cite{RDDVS23}, using both TDKS and RR-TDDFT, keeping all other aspects of the calculation the same as much as possible, i.e. the same functional, the same basis set, and approximately the same treatment of continuum states.  Consistently, the TDKS calculation yields the dipole-instability, while RR-TDDFT does not; this holds over a range of different field parameters and classes of functional approximations.  We provide an argument for how to understand how the instability observed in the TDKS calculation is onset in Sec.~\ref{sec:why}. Finally, we conclude in Sec.~\ref{sec:concs}. 

\section{Response-Reformulated TDDFT}
\label{sec:RR-TDDFT}
RR-TDDFT proceeds by propagating a set of ordinary differential equations (ODEs) for the coefficients of the time-dependent many-body state expanded in a basis of field-free many-body states $\{\Psi_m\}$,
%, in a manner formally analogous to time-dependent configuration interaction (TDCI). A key distinction,
with a key point being that these many-body states are never explicitly constructed~\cite{DBM24}. All quantities entering the ODEs
can be obtained from ground-state DFT and response TDDFT, and all properties directly related to the density (i.e. representable by one-body multiplicative operators) can in principle be exactly obtained. Like the traditional TDKS formalism, RR-TDDFT is formally exact and its foundation relies on the Runge--Gross theorem~\cite{RG84}, which proves a one-to-one mapping between the time-evolving density and a one-body multiplicative external potential. But unlike TDKS, the exchange-correlation terms for RR-TDDFT need to be evaluated only near ground-state densities, not on the fully non-equilibrium density, which makes the approach much more accurate when adiabatic functionals are used.

The time evolution of the expansion coefficients is governed by
\begin{equation}
i\,\dot{C}_m(t) = E_m\,C_m(t) + \sum_{n} V^{\mathrm{app}}_{mn}(t)\,C_n(t),
\label{eq:coeff-dyn}
\end{equation}
where the matrix elements of the applied potential are
\begin{equation}
V^{\mathrm{app}}_{mn}(t)
= \langle \Psi_m | V^{\mathrm{app}}(t) | \Psi_n \rangle
= \int d^{3}r\; v^{\mathrm{app}}(\mathbf{r},t)\,\rho_{mn}(\mathbf{r}).
\label{eq:vapp}
\end{equation}
For the common case of a spatially uniform electric field $\bm{\mathcal{E}}(t)$, the coupling between the coefficients, Eq.~(\ref{eq:vapp}), reduces to $V^{\mathrm{app}}_{mn}(t) = \bm{\mathcal{E}}(t)\cdot \mathbf d_{mn}$, with the dipole moment given by
\[
\mathbf d(t) = \sum_{n,m} C_n^*(t)\,C_m(t)\,\mathbf d_{nm},
\]
where $\mathbf d_{nm} = \langle \Psi_n | \hat{\mathbf r} | \Psi_m \rangle = \int d^3 r\, \mathbf r\, \rho_{nm}(\mathbf r)$.

The required inputs for RR-TDDFT are the energies $E_m$ (for $m=0$ obtained from ground-state DFT and for $m>0$ from linear-response TDDFT~\cite{PGG96,C95,C96,GPG00}), transition densities out of the ground state, $\rho_{0m}(\mathbf r) = \bra{\Psi_0}\hat{n}(\mathbf r)\ket{\Psi_m}$ (obtained from linear response TDDFT), transition densities between excited states, $\rho_{mn}(\mathbf r) = \bra{\Psi_m}\hat{n}(\mathbf r)\ket{\Psi_n}$ (obtained from quadratic response TDDFT~\cite{PF18,SVHA02}), and state densities $\rho_{mm}(\mathbf r)$ (for $m=0$ obtained from ground-state DFT and for $m>0$ from linear response TDDFT~\cite{F01,FA02,BM25}).
Observables are obtained, in principle~\cite{RG84}, from the time-dependent density
\begin{equation}
n(\mathbf r,t) = \sum_{n,m} C_n^*(t)\,C_m(t)\,\rho_{nm}(\mathbf r).
\label{eq:density}
\end{equation}
For any observable not directly related to the density  an additional 'observable functional' $O[n,\Psi(0)]$, is needed to extract it from the density and physical initial state. This is similar to the TDKS case, where observables are formally instead functionals of the density and KS initial state.

As mentioned earlier, the advantage of RR-TDDFT over TDKS is that the domain of the needed exchange-correlation functional is far closer to the domain in which they were derived: only ground-state and linear and quadratic response exchange-correlation functionals are needed, not the fully non-equilibrium $v\xc[n;\Psi(0),\Phi(0)](\br,t)$, even though fully non-equilibrium phenomena are being simulated.  

Making an adiabatic approximation for the latter in Eq.~(\ref{eq:tdks}) is clearly a much more drastic approximation when the system has evolved far from any ground-state than making the same approximation for the ground-state and response functionals, since the evaluation of the exchange-correlation term in Eq.~(\ref{eq:tdks}) is on a system far from the adiabatic regime. In other words, adiabatic TDDFT with Eq.~(\ref{eq:tdks}) approximates non-equilibrium (i.e. non ground-state) exchange-correlation effects with ground-state exchange-correlation, which can lead to serious errors when the system is far from a ground state. In contrast, the separation of the electronic structure and time-dynamics in the RR-TDDFT approach means that the exchange-correlation effects need only to be evaluated on systems close to the ground state.  
%The advantage of RR-TDDFT over traditional time-dependent Kohn--Sham (TDKS) propagation becomes evident when considering the role of the xc functional. In TDKS, the orbitals evolve according to
%\ben
%i\,\partial_t \phi_i(\mathbf r,t) = \left[-\tfrac{1}{2}\nabla^2 + v_s(\mathbf r,t)\right]\phi_i(\mathbf r,t),
%\een
%with $v_s = v_{\mathrm{ext}} + v_H[n] + v_{xc}[n,\Psi_0,\Phi_0]$, requiring the xc potential in a fully non-equilibrium regime. In practice, this is typically approximated by an adiabatic form, $v_{xc}^{\mathrm A}[n](\mathbf r,t) = v_{xc}^{\mathrm{g.s.}}[n(t)](\mathbf r)$, which neglects memory effects and applies a ground-state functional outside its domain of validity.
As a consequence, as demonstrated in Ref.~\cite{DBM24}, RR-TDDFT successfully captures phenomena such as Rabi oscillations and charge-transfer dynamics using adiabatic functionals, even in cases where the same approximations fail qualitatively within the TDKS framework. The formalism allows TDDFT to be used with as much reliability and accuracy for non-equilibrium dynamics as it is used in the response regime for spectra. Equally, it is limited by the accuracy of the functional in the response regime, but since the development of improved functionals has been much more successful for response than for non-equilibrium regime, RR-TDDFT can take advantage of these improvements to describe non-equilibrium dynamics (e.g. when significantly involving double excitations~\cite{RBDM26}).

We note that a similar method that uses TDDFT response quantities in a time-dependent configuration-like framework has been applied to studying electron dynamics in strong fields by a number of groups~\cite{SS11b,MLR14b,HPT17,CL21}, and most recently it has been implemented in an efficient GPU-accelerated code to study dynamics in a large organic molecule (120 atoms)~\cite{KL26}. In those works, only linear response TDDFT quantities and auxiliary wavefunctions were used for the couplings (so relaxation effects for excited-state couplings are missing), and some of these works make a further Tamm-Dancoff-like approximation: RR-TDDFT puts these approximate approaches on a rigorous foundation. 

In this paper, we use RR-TDDFT to demonstrate that the unexpected dipole instability observed in the TDKS simulations of Refs.~\cite{RDDSV23,RDDVS23,HDDVRS23,HDDRS25,DRDHS26} is a consequence of the exchange-correlation approximation used, and that it vanishes when the same approximation is used within the RR-TDDFT framework.

\section{N$_2$ dynamics: TDKS vs RR-TDDFT}\label{N2_dynamics}
The simulations of Refs.~\cite{RDDSV23,RDDVS23,HDDVRS23,HDDRS25,DRDHS26} took place in real-space grid codes QDD~\cite{QDD} and EDAMAME~\cite{EDAMAME}. For RR-TDDFT we need quadratic response functionality for the computation of the excited-to-excited state couplings. 
While linear response TDDFT is available in standard codes, fewer have
quadratic response capability, and  we are not aware of such a real-space code. We use the finite basis set code turbomole~\cite{turbomole} to extract the energies and couplings for RR-TDDFT, and compare with the real-time TDKS evolution performed with NWChem~\cite{nwchem} in the same basis set to make a consistent comparison. First we discuss some computational considerations.

\subsection{Computational Details}
\subsubsection{Basis set}
We will use the Gaussian basis set cc-pVDZ in all our calculations, except when otherwise noted. %which, for N$_2$, gives a basis set of size 14 spatial atomic orbitals on each atom.
The number of many-body states used in the RR-TDDFT calculations shown is 70, however the results are converged with only 40 states for all the cases we consider (this will be explicitly shown later). The  excited-to-excited state couplings involve a quadratic response calculation within which the pseudowavefunction approximation was applied to prevent any spurious divergences~\cite{PRF16,LL14, OBFS15,ZH15,PRF18}.

\subsubsection{Complex Absorbing Potential}
A key consideration in both the real-space simulation of
Ref.~\cite{RDDVS23} or our finite-basis set calculation, is the treatment of the ionization continuum: in reality, the part of the electronic wavepacket excited above the ionization threshold will propagate outwards, leaving the system unless driven  back by the electric field. To model this ionization, and prevent artificial reflections from boundaries, the real-space grid is typically outfitted with either imaginary potentials or a mask function at the boundaries~\cite{RSAMS06,RDDVS23}. 
With a finite basis-set code, this is more challenging: some approaches use a  real-space absorbing potential around each atom~\cite{KSS14,SAMGSL16,DS25} requiring a large number of diffuse functions in the basis to connect with them, or plane-wave complements are used~\cite{MLR12}.  
A simpler approach was introduced in Ref.~\cite{LG13b} where a damping coefficient on molecular orbital energies lying above a threshold is used to mimic the effect of an absorbing potential, and this is implemented in the NWChem software package. 
%which implements the CAP by constructing a diagonal damping matrix, $\Lambda$, within the molecular orbital basis. Each diagonal element of this $\Lambda$ matrix corresponds to an energy-dependent decay rate, $\gamma_i$, assigned to orbital $i$. 
Essentially, for the $i$th orbital energy above a fixed threshold, $\epsilon_i \to \epsilon_i - i \gamma_i$  where the 
damping coefficient is defined as:
\begin{equation}
\gamma_i =
%\begin{cases}
%0, & \epsilon_i \leq E_0 \\[6pt]
\gamma_0 \left[ \exp\left(  \xi (\epsilon_i - \epsilon_{th}) \right) - 1 \right],  \epsilon_i > \epsilon_{th}
%\end{cases}
\label{eq:gamma_nwchem}
\end{equation}
where $\gamma_0$ is the global damping rate scaling factor, and $\xi$ controls the steepness of the effective complex absorbing potential (CAP)~\cite{LG13b}. 
We set the CAP parameters as: $\gamma_0 =1$H, $\xi = 0.5$H$^{-1}$ with 
the threshold energy $\epsilon_{th}$ set to zero, and we clamp the upper limit of $\gamma_i$ to be 100 H. 
While we do not expect the results from molecular orbital damping applied in this way to correspond exactly to the real-space absorption by a mask or complex boundary, we do expect the effect is similar~\cite{LG13b}.

%At each timestep, the damping matrix $\Lambda$ is projected onto a matrix consisting of the instantaneous MO eigenvectors. This term is then added as an imaginary contribution to the time-dependent Kohn–Sham Hamiltonian. This renders the propagation non-Hermitian, ensuring that amplitudes associated with high-energy orbitals are exponentially attenuated during the time evolution~\cite{LN11}.

To emulate the effect of the CAP within the RR-TDDFT framework, an analogous damping term can instead be introduced at the level of the state amplitudes. In this case, the equations of motion are modified by appending a term of the form $-i\Gamma_m C_m$, on the RHS of Eq.~(\ref{eq:coeff-dyn}) for states with energies above the ionization threshold, in a similar spirit to Eq.~(\ref{eq:gamma_nwchem}). However, the parameters used in Eq.~(\ref{eq:gamma_nwchem}) would not in general translate over directly, since Eq.~(\ref{eq:gamma_nwchem}) applies to molecular orbitals, not to many-body states. For our case, we assume that each TDDFT excitation at an energy above the ionization threshold, is dominated by KS excitations with virtual orbitals at the same or near-lying energies so that using the same parameters could be justified. 
%{\color{violet}\it neepa checked this is true for the nwchem pbe/ccpVDZ response calc}
%To make a meaningful comparison, we optimize the exponential decay constant and prefactor to match the total ionization probability (loss of norm) from the corresponding TDKS calculation. 
%While this is not a unique fix, it is justified for our purposes of comparing RR-TDDFT with TDKS, especially since this quantity is directly related to the density, which, were the exact functional to be used in both schemes, would be identical.

\subsubsection{Linear response spectrum and XUV pulse considerations}
\begin{figure}[h]
    \centering
    \includegraphics[width=1\linewidth]{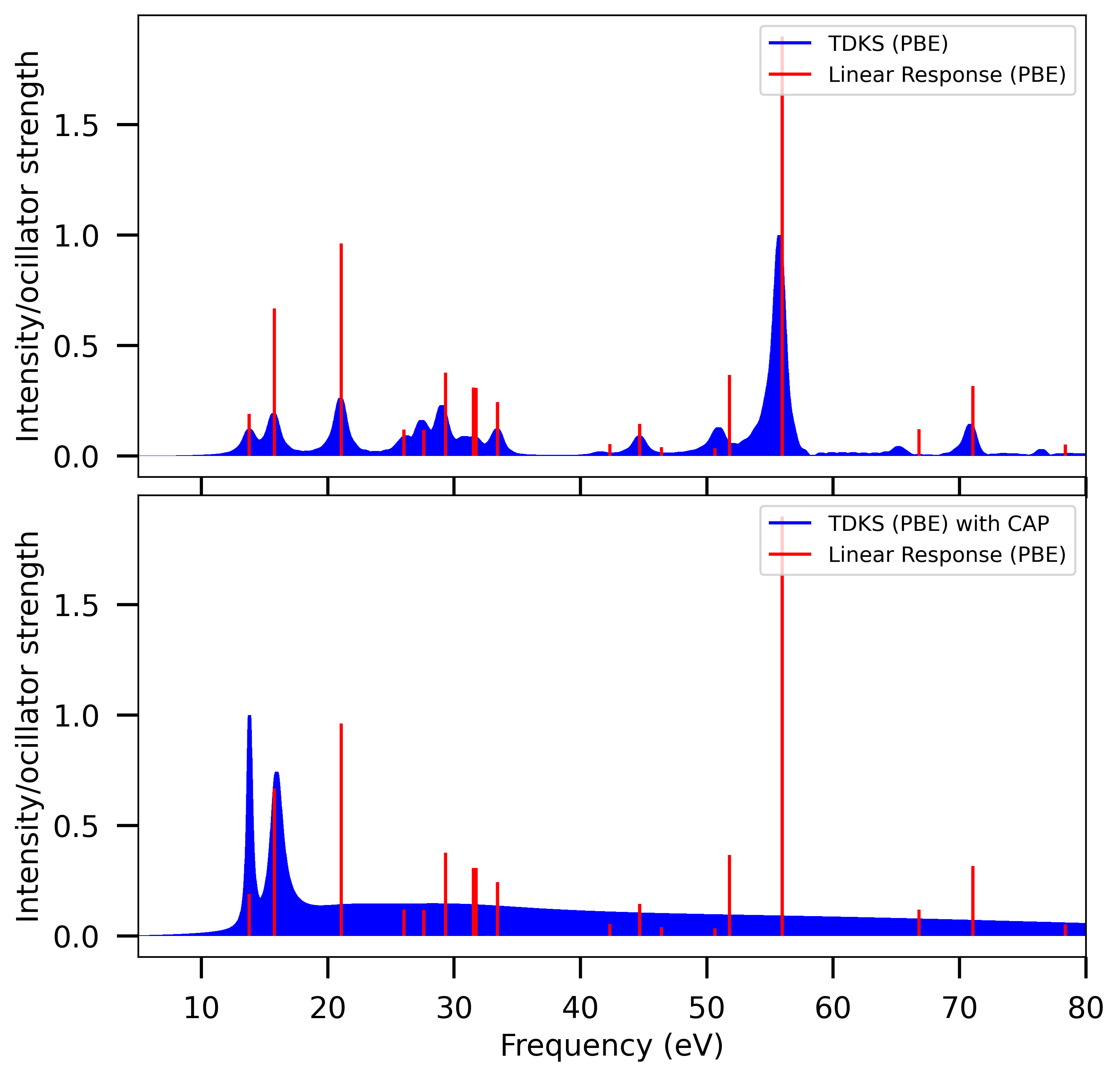}
    \caption{
    Comparison of TDDFT (PBE) absorption spectra for N$_2$ from linear response equations (red stick spectra) and resulting from a kick applied to the TDKS equations (blue  filled-in curve); the blue curve is the real-time TDKS intensity normalized to unit maximum,
$I_{\rm RT}^{\rm norm}(\omega)=I_{\rm RT}(\omega)/\max_{\omega} I_{\rm RT}(\omega)$,
while the red sticks denote the linear-response TDDFT oscillator strengths.  The upper panel shows the spectrum obtained from the kick without the CAP applied.  The lower panel shows the TDKS spectrum with CAP, with the same linear-response calculation shown for reference. 
 }
    \label{fig:spectra}
\end{figure}

Since the details of the electronic structure differ from that used in the real-space grid studies of Refs.~\cite{RDDVS23},   the same pulse may lead to different dynamics. Neither the local density approximation with average-density self-interaction correction (ADSIC) functional used in that work, nor the relaxation-time approximation (RTA) also explored there, are  available in the two codes we are using. However, we do not believe the choice of functional is a critical aspect of the phenomenon, as our examples will also explicitly show. Further, Ref.~\cite{RDDVS23} utilized a pseudopotential freezing the cores, while our calculations are all-electron. 
As much of our analysis will focus on the PBE functional~\cite{PBE96} in the cc-pVDZ basis, we show the corresponding spectrum in Figure~\ref{fig:spectra}. The stick spectra in both panels indicate the results from frequency-domain matrix equations of linear response, computed in turbomole (essentially the same results as NWChem) with positions at the excitation frequencies and heights determined by the oscillator strength. The  smoother filled-in curve shows the results obtained from a real-time calculation of the response, resulting from Fourier-transforming the dipole after a weak $\delta$-kick~\cite{YNIB06}; no CAP is applied in the top figure which shows good agreement with the stick spectrum from the frequency-domain calculation, with likely increasing agreement and less broadening with increasing duration of the time over which the Fourier transform is taken. 
The bottom figure applies the CAP, showing the dampening of the oscillator strength above the ionization threshold (computed from the PBE HOMO orbital energy, 9.76 eV). 

As expected the lower panel differs in detail from the ADSIC spectrum computed in the real-space basis with absorbing boundaries shown in Fig 1a of Ref.~\cite{RDDVS23}, but the trends are similar. In particular, we note that the XUV frequency of $\omega = 58.5$eV that was chosen to illustrate the dipole instability in Ref.~\cite{RDDVS23}, has the same property of being off-resonant ({\it c.f.} Figure 1a and 1d of Ref.~\cite{RDDVS23}).
It may be useful to note  that this frequency lies between the 70th and 71st TDDFT excitation for PBE/cc-pVDZ. Our cases will focus on this frequency as well.

As in Ref.~\cite{RDDVS23}, we will apply an ultrafast XUV pulse with the electric field
\ben
\mathcal{E}(t) =  \mathcal{E}_0\sin^2\left(\dfrac{\pi t}{T_{\rm pulse}}\right)\sin(\omega_c t), \;\;\; 0<t<T_{\rm pulse}
\label{eq:pulse}
\een
such that $v^{\rm app}(\br,t) = \mathcal{E}(t)z $ (atomic units are used throughout unless otherwise stated). 
Ref.~\cite{RDDVS23} chose pulses of duration of 1fs centered in the XUV range, with an intensity such that about 1 electron was absorbed. 
We will return to the selection of pulse parameters for our study in Sec.~\ref{sec:instability}.

For both the TDKS calculation run in NWChem~\cite{nwchem} as well as the RR-TDDFT coefficient evolution run in our in-house code, we use a time step of 0.1 a.u.

%In both real-space grid and finite-basis set calculations, this high-energy excitation typically leads to unphysical reflections at the simulation boundaries. To mitigate these artifacts, the real-space calculations in Ref.~\cite{RDDVS23} utilize a complex absorbing potential (CAP). The CAP is implemented by damping the time-dependent molecular orbital (MO) coefficients associated with high-energy virtual orbitals. This approach causes the continuum components of each orbital to decay during temporal propagation while leaving the bound components largely unaffected.In the present work, we employ a CAP within our Time-Dependent Kohn-Sham (TDKS) framework. 

\subsubsection{Computational cost considerations}
First, we note that while we do not have a numerically exact electronic structure reference to compare our TDDFT methods against, we will use the approximate coupled-cluster method CC2 as an alternative wavefunction-method reference~\cite{CKJ95,HHYJ00, SSST08}, which has often been used to benchmark TDDFT excitation energies~\cite{LJ13,SSST08}. CC2 is an approximation to CCSD, in which singles-dominated excitations are known to well-approximate the CCSD result, while double-excitations are correct only to zeroth order.

Excitation energies and interstate dipole couplings are computed for both CC2 and RR-TDDFT in turbomole, 
in the cc-pVDZ basis set, with the cost/time for calculation of couplings between excited states from quadratic response scaling far outweighing the linear response. 
For 40 excited states and their interstate couplings, CC2  required approximately 115 CPU hours. In contrast, the TDDFT/PBE required only 4 CPU hours.
This highlights the substantially lower computational scaling of TDDFT-based approaches relative to CC2, particularly as the number of states and couplings increases.
In our TDDFT calculations we found that our results were well-converged with 40 states, showing little difference when 80 states were included; going up to 70 states and couplings using TDDFT increased the computational time to 35 CPU hours. 
%{\color{purple} \it maybe a bit weird since 58.5eV lies between the 70th and 71st TDDFT excitation...i have to think about that,i guess the field is strong enough to be primarily multi-photon dynamics but then it would suggests doubles important? will need some discussion somewhere, maybe not right here}

The computational cost/timing of the coefficient evolution for RR-TDDFT or the TDCI/CC2 is negligible compared to the time to obtain the interstate couplings. It is important to note that the expensive part of the calculation needs to be done only once and for all for a given basis set and functional choice: for the application to dynamics under any field, these same couplings are input and this part of the calculation (propagation of Eqs.~(\ref{eq:coeff-dyn})) takes a few seconds, involving solving only a system of $M$ ODEs where $M$ is the number of many-body states. In contrast, the TDKS calculation requires solving $N$ non-linear partial differential equations in space and time, with a different calculation for each applied field.

\subsection{N$_2$ Dipole Instability}
\label{sec:instability}
\begin{figure}[h]
    \centering
    \includegraphics[width=\linewidth]{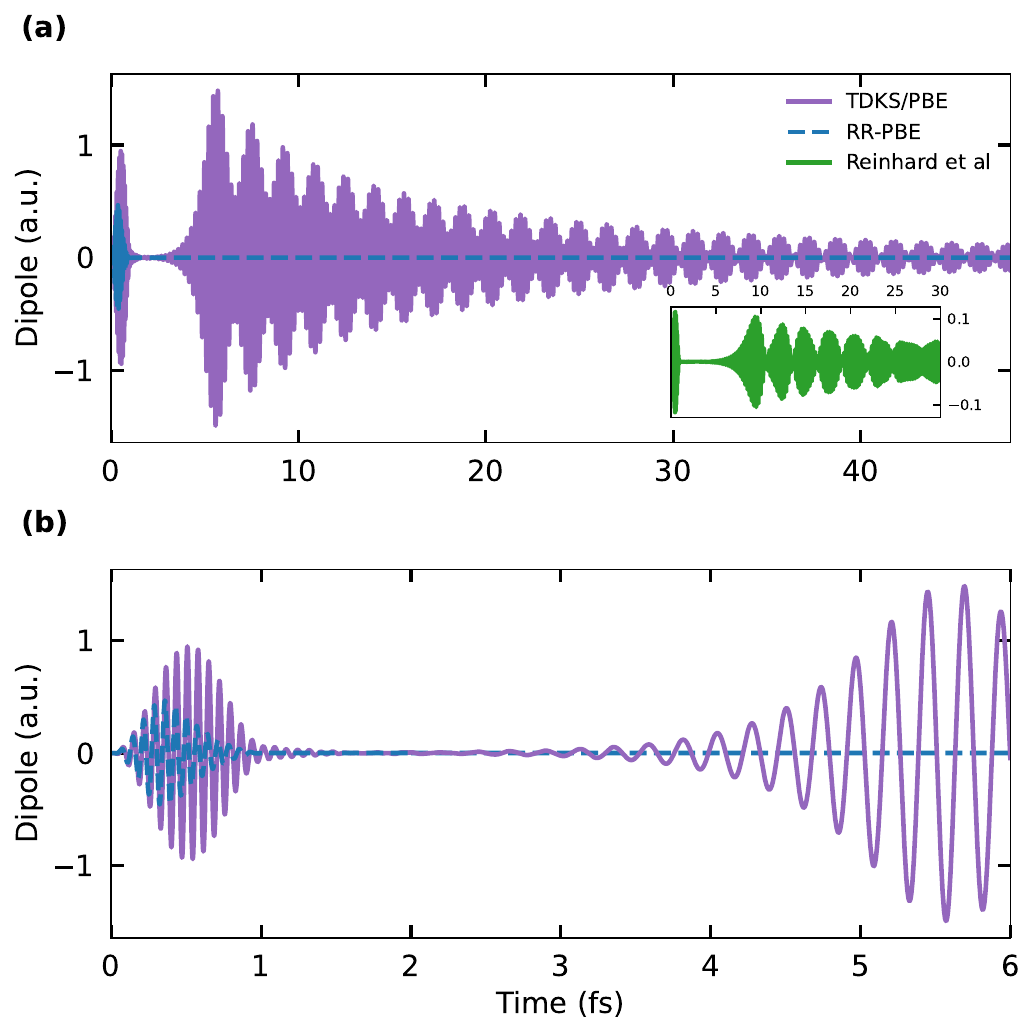}
    \caption{
     Time-dependent dipole of N$_2$ computed using TDKS and RR-TDDFT approaches under the applied field ($E_0 = 0.45$ a.u., $\omega = 58.5$~eV, $T_{\rm pulse}=1$~fs in Eq.~\ref{eq:pulse}), both within PBE/cc-pVDZ. \textbf{Top panel:}  TDKS (purple, solid) and RR-TDDFT (blue, dashed) predictions using  70 states. Couplings between excited states in RR-TDDFT were computed within the pseudowavefunction approximation, and both calculations used the effective CAPs described in the text.
     The inset shows the results from Ref.~\cite{RDDVS23}, computed from TDKS within LDA-ADSIC in a real-space grid with a mask function absorbing boundary.   \textbf{Bottom panel:} As in the top panel but zoomed into the early-time regime (0--6~fs). 
    }
    \label{fig:w58eVE01fs}
\end{figure}

We begin by comparing the results of the pulse that was used in Fig. 1d of Ref.~\cite{RDDVS23}. This is:
$E_0 = 0.45$~a.u. $= 7 \times 10^{15}$ W/cm$^2$, $\omega = 58.5$~eV, and $T_{\rm pulse} = 1$~fs.
Figure ~\ref{fig:w58eVE01fs} shows that the dipole-instability is dramatically evident in our finite-basis TDKS calculation (PBE/cc-pVDZ), occurring in about the same timescale as that in Ref.~\cite{RDDVS23} whose dipole is reproduced in the inset of the top panel. Due to the different functional, basis set, and CAP details, the details are different, and the dipole we obtain is about an order of magnitude greater than that in Ref.~\cite{RDDVS23}. The number of electrons ionized during the pulse is about 1.3, greater than the one electron ionized in Ref.~\cite{RDDVS23}. 

Turning now to the RR-TDDFT calculation run under the same conditions as our TDKS simulation, we find that the instability is completely absent in RR-TDDFT (shown here including 70 states in the evolution of the coefficients in Eq.~(\ref{eq:coeff-dyn}) but the results are converged using 60 states). 
This may not be surprising: from the linear structure of the RR-TDDFT equations Eqs.~\ref{eq:coeff-dyn}, it would be difficult to imagine how the approach would give any change in the magnitude of the dipole oscillations in the absence of any applied field. 

Zooming in on earlier times in the lower panel  shows that TDKS agrees with RR-TDDFT for the first few cycles, and then differs: the amplitude of the oscillations in TDKS continues to grow until the peak of the applied pulse, while RR-TDDFT peaks earlier and the asymmetry in the fall-off is more significant than in TDKS, with more ionization. We believe this is an inherent difference due to the two different approaches~\cite{DBM24,RBDM26} (see also Sec.~\ref{sec:why}),  not a consequence of any choice of numerical parameters.
Any limitation in the basis set size affects both calculations.  
The dipole computed within the larger aug-cc-pVDZ basis set shows only a small difference for the RR-TDDFT and TDKS calculations during the pulse,  and the instability in the latter is onset at a slightly larger time but has the same shape and features otherwise (not shown here). 
However, for the purposes of investigating the dipole instability, these pulse parameters are not ideal for our numerical set-up, since after the pulse is off the oscillations of the dipole never fall below about 0.01 a.u, which is two orders of magnitude larger than the oscillations left in the dipole of Ref.~\cite{RDDVS23}.

\begin{figure}[h]
    \centering
    \includegraphics[width=\linewidth]{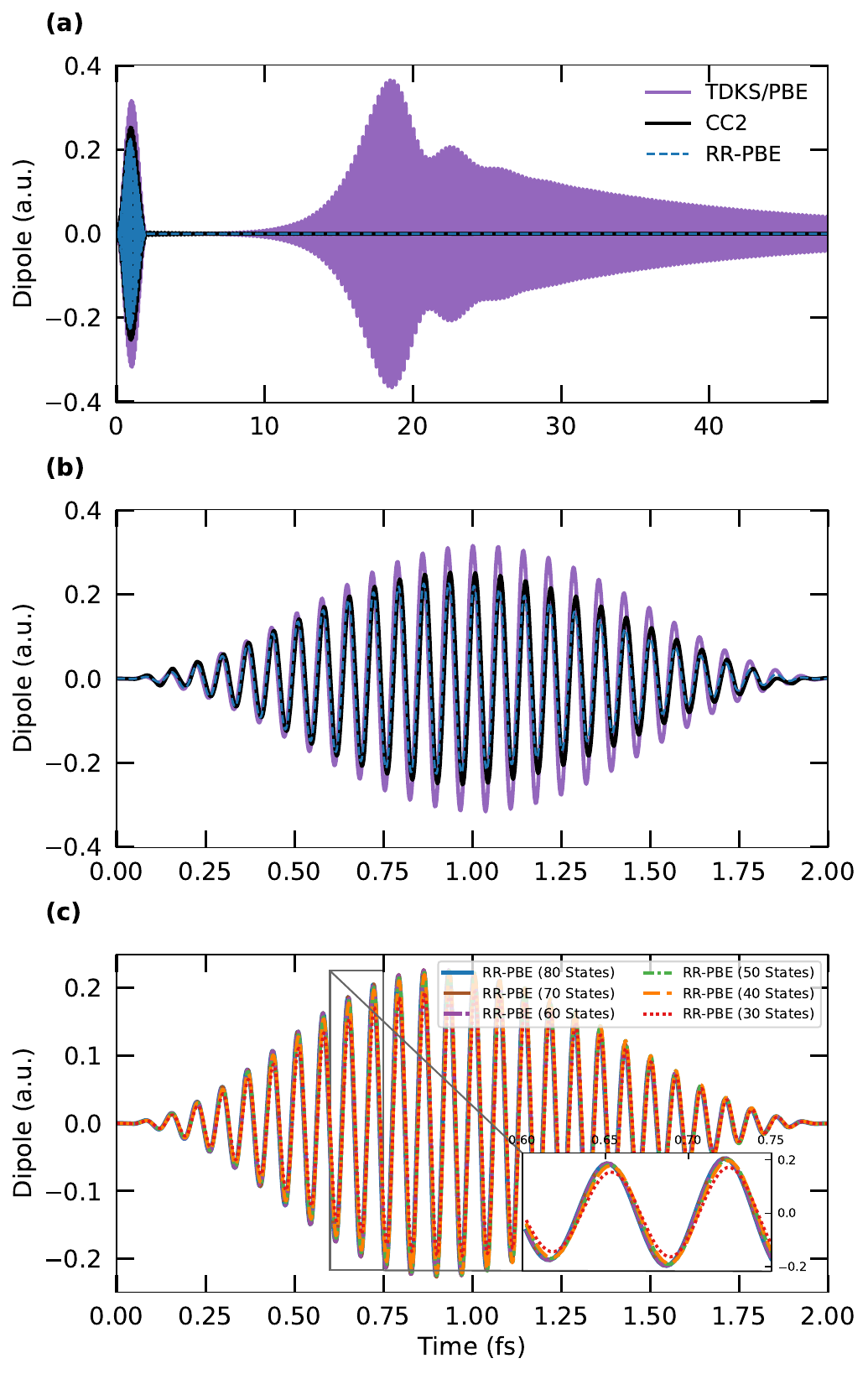}
    \caption{ 
    Time-dependent dipole dynamics computed using time-dependent TDKS, RR-TDDFT, and CC2 approaches under the applied field ($E_0 = 0.15$ a.u., $\omega = 58.5$~eV, $T_{\rm pulse} = 2$~fs). \textbf{Top panel:} 
    TDKS and  RR-TDDFT  dipoles with the PBE functional (cc-pVDZ) and CC2 using 40 states. \textbf{Middle panel:} The dipoles shown during the pulse. All simulations include the CAP described in the text.  \textbf{Bottom panel:} Convergence of the RR-TDDFT dynamics as a function of the number of states. } 
    \label{fig:w58.5E0ov32fs}
\end{figure}

We therefore focus on a weaker pulse with parameters $E_0 = 0.15$~a.u., $\omega = 58.5$~eV, and $T_{\rm pulse} = 2$~fs.  
Figure~\ref{fig:w58.5E0ov32fs} shows the dipole resulting from the TDKS and RR-TDDFT calculations, both again using the PBE functional in the cc-pVDZ basis.   After a longer quiet initial post-pulse behavior, the dipole-instability in the TDKS evolution is again evident, and completely absent in the RR-TDDFT approach. The middle panel shows that
during the pulse, the two methods give similar results, 
practically matching during the first five or so cycles, and remaining in phase throughout, with TDKS climbing to larger oscillation amplitudes towards the second half of the pulse.   With the chosen CAP 0.38 electrons are ionized after the pulse in the TDKS calculation before the onset of the dipole resurgence, while 4.5 are ionized in the RR-TDDFT calculation.
The RR-TDDFT dipole matches very closely with the TD-CI approach using CC2 energies and couplings as input, both during the pulse and in their absence of instability; using the same CAP function as in the RR-TDDFT calculation produces an ionization yield of 1.84 electrons in the CC2 calculation. 
%{\color{purple} \it dhyey, right? maybe we need to footnote comment about how we dont have the CC2 ionization energy..or not...}. 
Again, we believe the salient difference observed between the TDKS and RR-TDDFT calculations (different ionization yields and, especially, instability behavior) is inherent to the methods themselves rather than being a consequence of numerical parameters. At the same time, it should be noted that to understand quantitative differences in the dynamics during the pulse and the ionization yields, a deeper investigation of what the equivalent CAP should be between the two methods would be needed. 

\begin{figure}[tb]
    \centering
    \includegraphics[width=\linewidth]{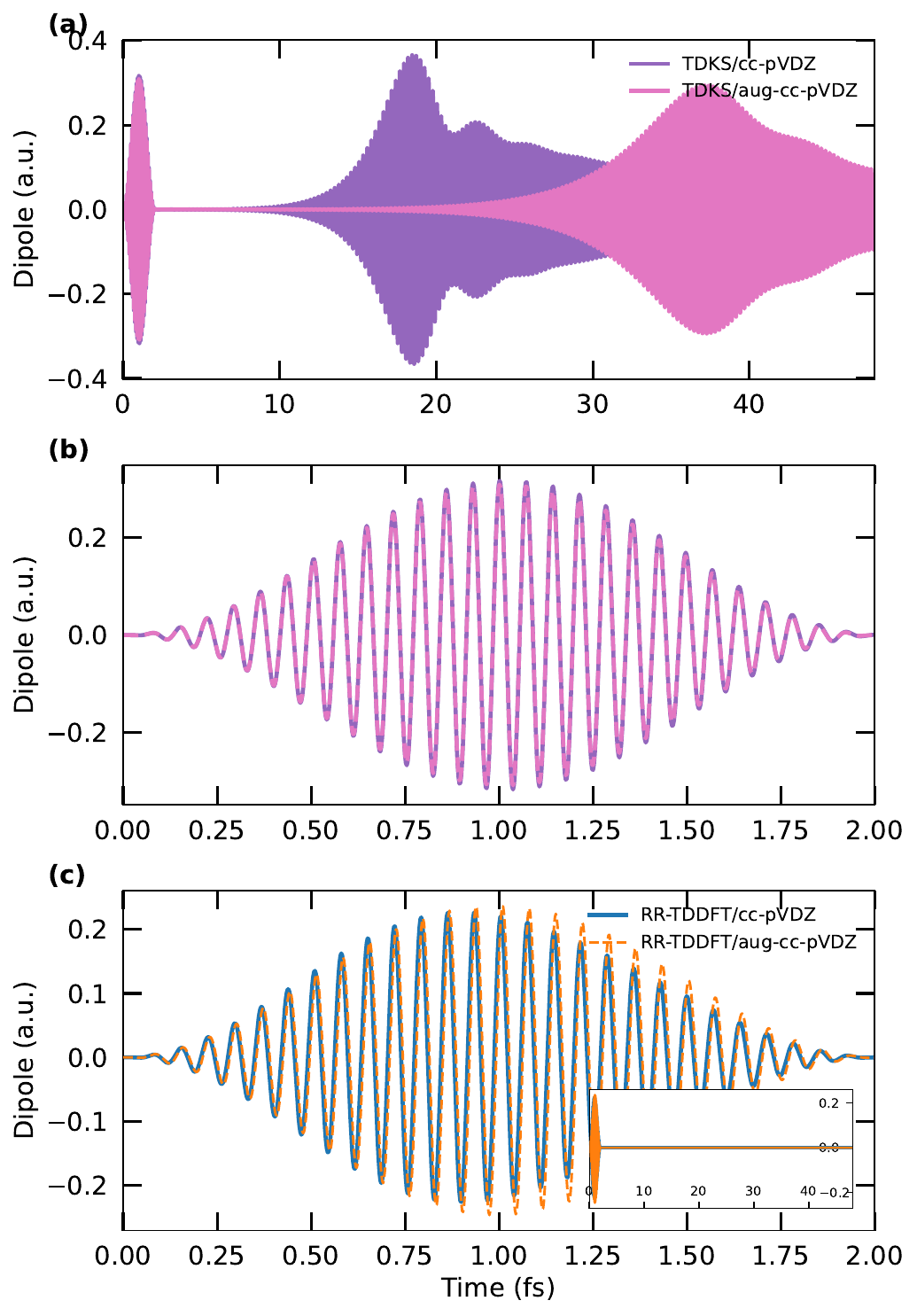}
    \caption{Time-dependent dipole dynamics illustrating basis set dependence on TDKS 
and RR-TDDFT approaches under the applied field ($E_0 = 0.45$~a.u., 
$\omega = 58.5$~eV, $T_{\rm pulse} = 2$~fs). \textbf{Top panel:} TDKS/PBE with cc-pVDZ and aug-cc-pVDZ basis sets. 
\textbf{Middle panel:} As in top panel but zoomed in for times during the pulse. \textbf{Bottom panel:} RR-TDDFT/PBE with energies and couplings computed in cc-pVDZ and aug-cc-pVDZ as indicated. 
The inset shows the 
corresponding evolution over a longer time range. }
    \label{fig:augccpvdz_vs_ccpvdz}
\end{figure}

The bottom panel of Fig.~\ref{fig:w58.5E0ov32fs} shows the RR-TDDFT dipole computed using different numbers of states and their couplings, showing that the dipole converges with about 40 states included.  
To confirm that the basis set choice is not a critical factor in the phenomenon, we compare the dipoles using the aug-cc-pVDZ basis in Figure~\ref{fig:augccpvdz_vs_ccpvdz}. The top panel shows that the instability remains in the larger basis set, albeit delayed and we note that the ionization yield reduces slightly to 0.368 electrons during the pulse.
The middle panel shows that during the pulse there is almost no difference in the dipole dynamics. Again, RR-TDDFT shows no instability, as shown in the bottom panel, and there is a small difference in the oscillations during the pulse, with the  ionization yield after the pulse reducing to 2.78 electrons.

\begin{figure}[h]
    \centering
    \includegraphics[width=\linewidth]{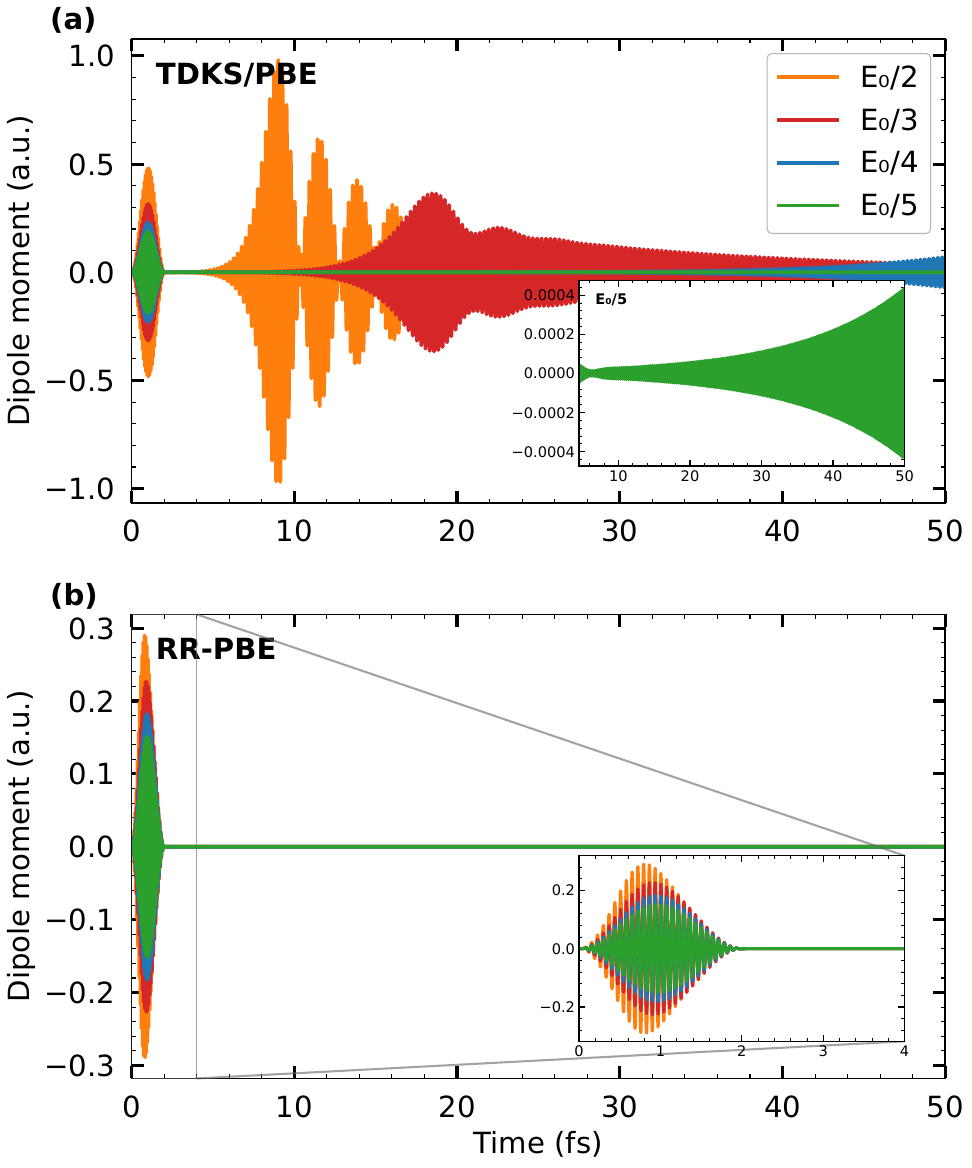}
    \caption{
    Time-dependent dipole moment under the 2fs, $\omega_c = 58.5$eV pulse with varying pulse strengths: $E_0/2$ (orange), $E_0/3$ (red), 
    $E_0/4$ (blue), and $E_0/5$ (green), where $E_0 = 0.45\,\mathrm{a.u.}$:
    (a) TDKS/PBE and  
    (b) RR-PBE.
    }
    \label{fig:energy_comp}
\end{figure}

Figure~\ref{fig:energy_comp} shows the TDKS and RR-TDDFT dipole dynamics for a range of field-strengths. The larger the field strength, the faster and larger the onset of the dipole resurgence for this range of intensities; note that while on the scale of the initial dipole driven by the field, the instability is barely visible for the weakest field shown ($E_0/5$) but zooming in shows it is there (see inset in the upper panel). 

In all cases there is no growth whatsoever in any of the RR-TDDFT results; as noted earlier, the linear nature of Eqs.~(\ref{eq:coeff-dyn}) preclude such a growth in the absence of the field. 

\subsection{Towards understanding the dipole-instability of TDKS}
\label{sec:why}
\begin{figure}[h]
    \centering
    \includegraphics[width=\linewidth]{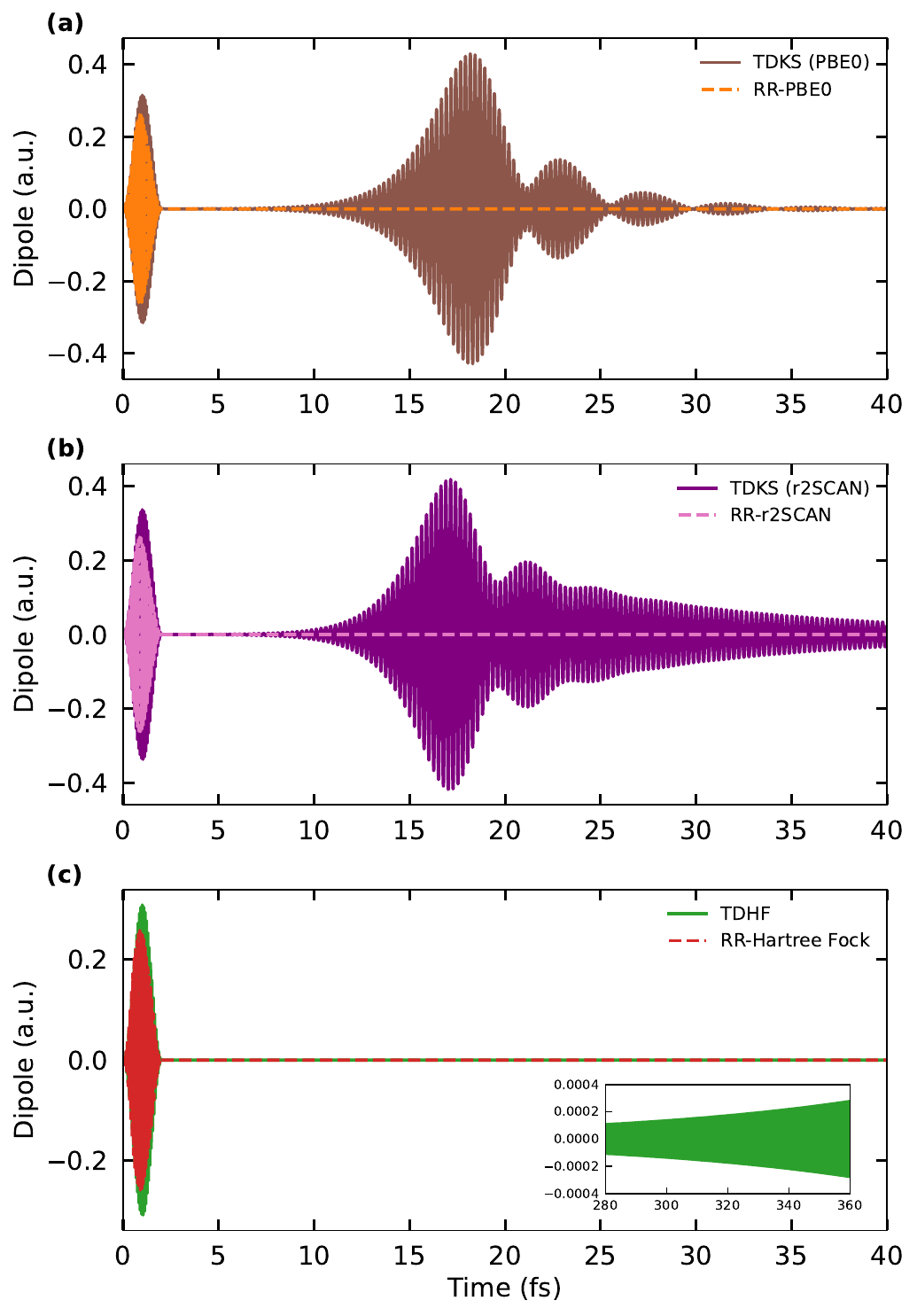}
    \caption{
    Dipole instability in TDKS under different functional approximations, compared against RR-TDDFT with the same functional. \textbf{Top panel:} Hybrid functional PBE0  \textbf{Middle panel:} Meta-GGA functional r$^2$SCAN .  \textbf{Bottom panel:} Hartree-Fock.   Note the different timescales in each of the panels.  }
%    While the PBE0 and Hartree-Fock are evolved under the same pulse as in Fig.~\ref{fig:w58.5E0ov32fs}, the Hartree TDKS evolution does not show instability for those parameters; shown in the bottom panel is instead the dipole for $E_0 = 0.225$ a.u.}.}
    \label{fig:manyfnals}
    \end{figure}

These results show that using the PBE functional within TDKS gives the dipole-instability while PBE used within RR-TDDFT does not. This is not a feature particular to the PBE exchange-correlation functional (nor the LDA-ADSIC or RTA explored in Refs.~\cite{RDDSV23,DRDHS26,HDDRS25,RDDVS23,HDDVRS23}), but applies to many classes of functional approximations. We show this in Figure~\ref{fig:manyfnals}, where in contrast to the GGA class of functional (PBE), we show a hybrid (PBE0)~\cite{PBE0}, a meta-GGA (r$^2$SCAN)~\cite{SRP15,r2SCAN}, and the self-interaction free correlation-less time-dependent Hartree-Fock (TDHF)~\cite{TDHF}. The onset and other details of the post-pulse dipole-instability can differ significantly depending on the functional. For example,
for TDHF, the onset is much slower such that the same oscillation amplitudes are reached hundreds of femtoseconds later than for PBE, PBE0, or r$^2$SCAN. For a more intense pulse of amplitude three times that shown here,  the maximum of the first dipole resurgence is reached at around 175 fs in TDHF, but around 5fs in PBE. This implies that self-interaction may play a key role in the instability, but while a self-interaction-free functional delays the onset, it does not eliminate it. 
We also note the ionization yields during the pulses differ depending on the functional:  within TDKS the ionization yields were 0.338, 0.397, and 0.273 electrons for PBE0,  r$^2$SCAN, and TDHF respectively, while larger (4.9, 5.6, and 4.45) respectively for RR-TDDFT. Although not shown here, if we adjust the CAP so as to get a similar ionization yield from RR-TDDFT than for the TDKS, RR-TDDFT still does not produce the dipole-instability. 
As noted in Refs.~\cite{RDDVS23, HDDRS25}, the appearance or absence of the instability also depends on the frequency of the pulse; this is true also for the finite-basis case (although again not shown here).

In all cases, there is no dipole-instability in RR-TDDFT: In the absence of a field, only the free-evolution term of Eqs.~(2) (first term) is active, and the magnitudes of the coefficients remain the same as they were when the field was turned off except for the exponential decay the CAP causes for states above the ionization threshold. That is, Eqs.~(2) reduce to $
 i\dot C_m(t)=\bigl(E_m-i\Gamma_m\bigr)C_m(t)
$, where $\Gamma_m$ is the damping parameter for the $m$th state (Sec III.A.1), so that the time-dependent dipole is  
\begin{align}
 d(t)=\sum_{mn} &C_n^*(T)C_m(T)d_{nm}
 \exp[-i(E_m-E_n)(t-T)] \nonumber\\
 &\times \exp[-(\Gamma_m+\Gamma_n)(t-T)].
\end{align}
where $T = T_{\rm pulse}$ is the time the pulse is turned off.
That is, the dipole predicted by RR-TDDFT may display dampened beatings among coherences that had been created by the pulse, but  an exponential type of growth for a finite period of time is simply not allowed by this mathematical form.  The linear structure of the ODEs ensures not only a stable propagation but also makes it impossible for such a growth to occur without a field present.  While the separation of spatial- and time-dependence was noted to be advantageous from the point of view of exchange-correlation functional dependence in Ref.~[27], here we see another advantage is the removal of unphysical behavior arising from incorrect non-linearity (see more shortly). 
We note that the molecule returns primarily back to the ground-state after the pulse; due to the off-resonant nature, even during the pulse only a  small fraction (maximum 2.54\%) of the electron density enters the excited states as shown in  Fig.~7.

We note that the dynamical part of the equations of RR-TDDFT is the same as for the exact many-body dynamics: the approximations of RR-TDDFT enter only in static ingredients (energies, densities, and transition densities) that are input in the evolution equations and, in practise, in the truncation to a finite expansion basis. If we were to instead use exact static many-body wavefunctions to evaluate these ingredients, we could not get an instability for the same argument as above. In contrast, there is no analog for the incorrect non-linearity of the adiabatic TDKS Hamiltonian  that we will now show leads to the instability in the many-body TDSE approach.

%The linear structure of the ODEs ensures not only a stable propagation but also makes it impossible for exponential growth to occur without a field present.  While the separation of spatial- and time-dependence was noted to be advantageous from the point of view of exchange-correlation functional dependence in Ref.~\cite{DBM24}, here we see another advantage is the removal of unphysical behavior arising from incorrect non-linearity. 

\begin{figure}[h]
    \centering
    \includegraphics[width=\linewidth]{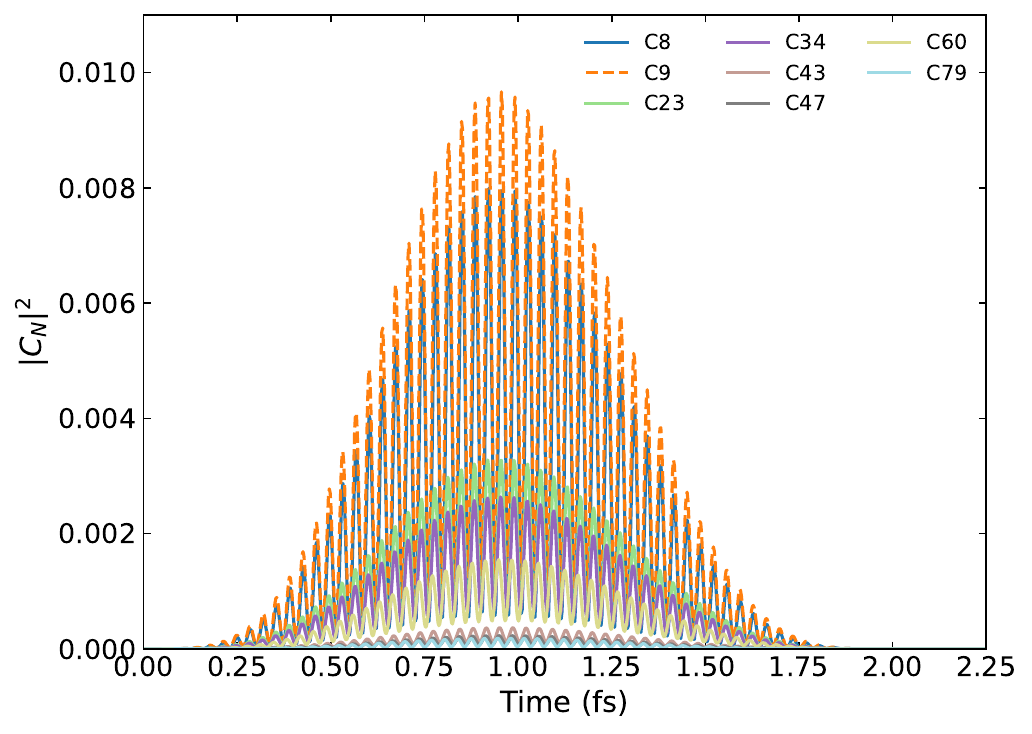}
    \caption{
    Time-dependent evolution of the largest excited-state populations ($|C_N|^2$) obtained from the RR-TDDFT simulation under an applied field ($E_0 = 0.15$, $\omega = 58.5$~eV, $T_{\rm pulse} = 2$~fs). Only states with a maximum population exceeding $10^{-3}$ are shown. The population dynamics exhibit coherent buildup and decay within the pulse envelope, with a small subset of states dominating the excitation manifold.
    }
    \label{fig:pseudo_populations_mainpulse}
\end{figure}

\begin{figure}[h]
    \centering       \includegraphics[width=\linewidth]{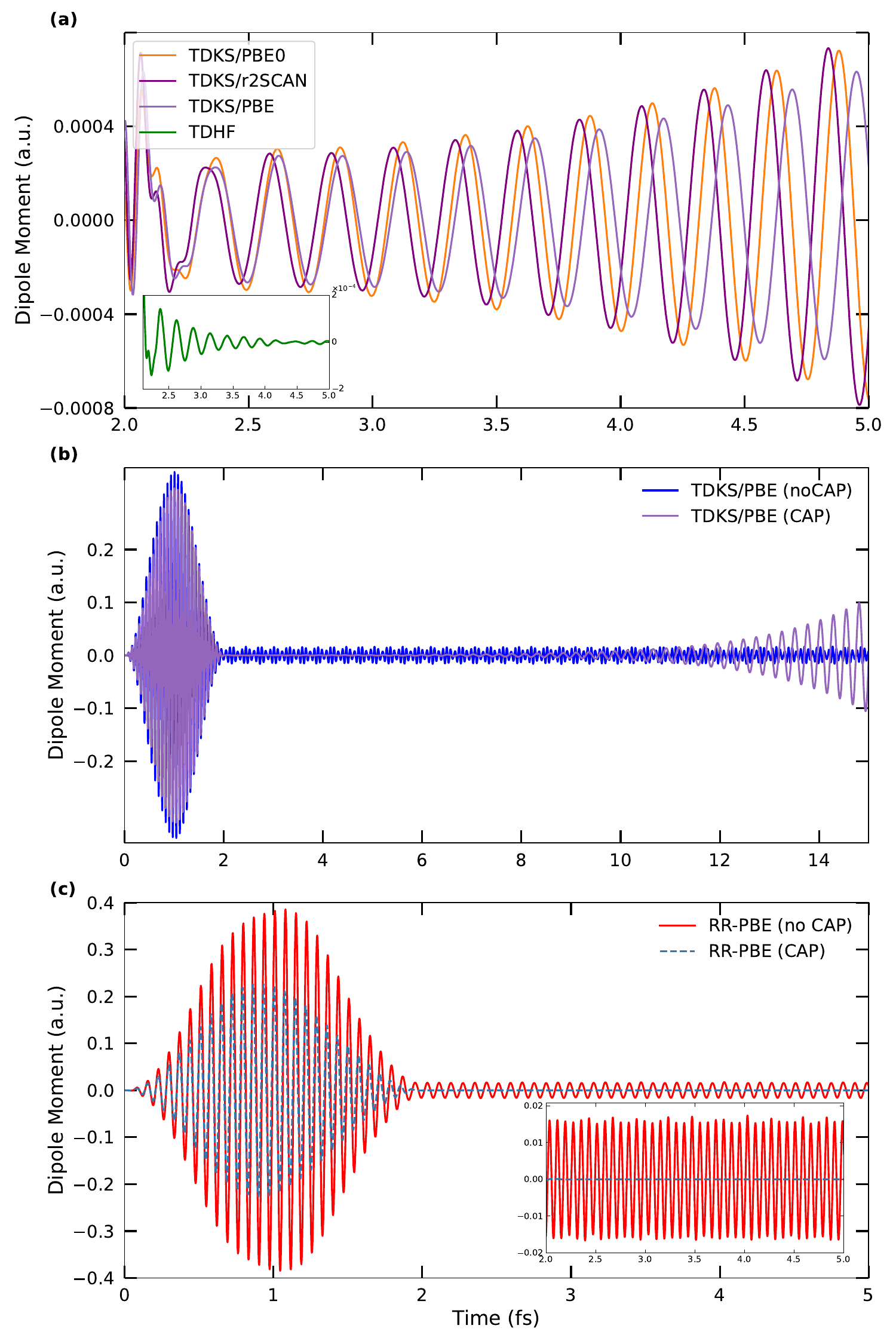}
\caption{
Time-dependent dipole dynamics computed using TDKS and RR-TDDFT approaches 
under the applied field ($E_0 = 0.15$~a.u., $\omega = 58.5$~eV, $T_{\rm pulse} = 2$~fs). 
\textbf{Top panel:} Zoomed view of the TDKS dipole evolution shortly after the 
pulse, with PBE, PBE0, r$^2$SCAN functionals, all with CAP 
applied. The inset shows the TDHF dipole over the same window.
The oscillation frequency observed in the 2--5~fs window for each 
functional corresponds to its respective 8th excitation frequency:
$0.0922$~a.u. (TDHF), $0.0922$~a.u. (PBE), $0.0922$~a.u. (PBE0), and
$0.0943$~a.u. (r$^2$SCAN).
The corresponding oscillation periods are approximately
$0.2624$~fs (TDHF), $0.2624$~fs (PBE), $0.2624$~fs (PBE0), and
$0.2565$~fs (r$^2$SCAN).
\textbf{Middle panel:} Comparison of the TDKS/PBE dipole with and without CAP 
applied \textbf{Bottom panel:} RR-TDDFT/PBE dipole dynamics with and 
without CAP during the pulse and shortly after. The inset shows a zoomed view between 2 and 5 fs.  
}    \label{fig:capvsnocap}
    \end{figure}

To try to understand how the dipole instability emerges in the TDKS system, consider the initial behavior of the dipole after the pulse is off. The top panel of Figure~\ref{fig:capvsnocap} shows that consists of very small oscillations of  frequency  about 16 eV in {\it all} cases where instability occurs, regardless of pulse parameters,  which corresponds to about the 8th excitation (the second noticeable peak in the spectrum) for each of the functionals. Given the off-resonant nature of the pulse and the coefficients in RR-TDDFT, we expect that even in TDKS, the system  returns largely to the ground-state after the pulse. We assert that
 during the initial period after the pulse is turned off, the density has the form: 
\ben
n(\br,t) = {\cal N}(n_0(\br) + \delta(t)\sin(\Omega_n t)n^{(1)}(\br) + O(\delta^2) )
\label{eq:density}
\een
where $\delta(t)$ is a small amplitude which varies in time more slowly than the frequency of oscillations $\Omega_n \approx 16eV \approx E_8 - E_0$, and $n_0(\br)$ indicates the ground-state density. For example, this form would arise if one (or more) of the TDKS orbitals after the pulse was left in the superposition $\phi_{\rm occ}(\br) + \delta(0)\phi_8(\br)$, such that its density is $\vert\phi_{\rm occ}(\br)\vert^2 + 2Re(\delta(0) \phi_{\rm occ}(\br)\phi_8(\br))$ to $O(\delta)$, in which case the 
normalization factor ${\cal N} = 1$ to linear order in $\delta$, because  the transition-density integrates to zero.
Evaluating the time-dependent Hartree-exchange-correlation potential, $v\Hxc = v\H + v\xc$, on the density of Eq.~(\ref{eq:density}) and expanding up to $O(\delta)$ we have:
\bea
\label{eq:argument}
v\Hxc^{\rm adia}[n](\br,t) &=& v\Hxc^{\rm adia}[n_0](\br) \\ &+& \underbrace{\int f\Hxc^{\rm adia}[n_0](\br,\br')n^{(1)}(\br') d^3r'}_{=\,C(\br)} \,.\,  \delta(t)\sin(\Omega_n t) 
\nonumber
\eea
This means that the TDKS Hamiltonian driving the orbitals is, to $O(\delta)$,
\ben
h\s = h\s^{(0)} + 2\delta(t) C(\br) \sin(\Omega_nt)
\label{eq:hs}
\een
That is, the TDKS system is driven by the field-free ground-state KS Hamiltonian plus a small driving term that is resonant with one of the linear-response frequencies of the  system. When any system is driven at its resonant frequency the oscillations grow, i.e. $\delta(t)$ will get larger until the perturbation approximation is no longer valid. But at that time, the larger changes in the density lead to the resonant frequencies of the TDKS system beginning to change significantly, as has been discussed in the literature as the spurious peak-shifting problem of adiabatic TDDFT~\cite{FHTR11,FLSM15,LFM16}, so that  the growth of the oscillations peaks and then falls. 

A key point in the above argument is the dependence of the functional in Eq.~(\ref{eq:argument})  on the instantaneous density. It is not the non-linearity {\it per se} of the functional, but rather the instantaneous dependence. 
The artificial self-driving effect would not arise with an appropriate memory-dependent functional (also nonlinear), as in the case of the exact functional, because such a functional would not be perfectly in sync with the density oscillations; the term $C(\br)$ in Eq.~(\ref{eq:argument}) would be time-dependent. The memory arising through the particular orbital-dependence of the hybrid PBE0, the meta-GGA r$^2$SCAN, or TDHF, appear not to be adequate to cure the effect.
The argument also rests on the assumption that the system is left with very small (pure or commensurate) oscillations after the pulse, and suggests that any time this happens there will be a tendency towards an exponential growth of these oscillations with an adiabatic functional i.e. the dipole-instability.  
The argument highlights the crucial role of the CAP:  without the CAP, the system is left oscillating with a wider, more noisy, spectrum of frequencies after the pulse, between which interference effects  prevent the self-driving effect to occur. Indeed, the dipole-instability vanishes in all cases when the CAP is turned off; an example is shown in Fig.~\ref{fig:capvsnocap}b. The CAP has less of a significant 
role in RR-TDDFT: other than the small post-pulse oscillations and larger signal during the pulse that occur when the CAP is removed,  the results are qualitatively similar with and without it, as shown in Fig.~\ref{fig:capvsnocap}c.
With the CAP on, we have found that in cases where the dipole-instability does not arise (not shown here), the system appears to be left oscillating in a more complex way with more frequencies. Our argument does not address {\it when} we can predict the system is left with essentially one (or a few commensurate) frequencies after the pulse, and a more complete investigation is left for future work.

%\begin{figure}[h]
%    \centering   %\includegraphics[width=\linewidth]{pbe_plot.pdf}
%    \caption{
%    Comparison of time-dependent dipole dynamics computed using RR-TDDFT with PBE0 and PBE exchange--correlation functionals under an applied field ($E_0 = 0.45$, $\omega = 58.5$~eV, $\T_pulse = 1$~fs). The main panel shows the early-time regime (0--2~a.u.), while the inset (bottom left) displays the full time evolution (0--48~a.u.). The two traces are indistinguishable on the plotted scale. All simulations include a complex absorbing potential (CAP). The CAP parameters were set to $\gamma = 1$, $\mathrm{IP} = 10.49~\mathrm{eV}$, and $\xi = 0.5$.
%    }
 %   \label{fig:pbe0_vs_pbe}
%\end{figure}

\section{Conclusions and Outlook}
\label{sec:concs}
While the adiabatic approximation in TDKS has produced useful results in many cases, and in particular has enabled the study of electron dynamics in systems too large to be studied otherwise, e.g. Refs.~\cite{DAGBSC17,
LGIDL20,SK18,S21,SZYYK21,BCV22}, the lack of memory-dependence has also been found to be the root of large errors, unreliable results, and failures~\cite{WU08,RN11,RN12c,RN12b,HTPI14,QSAC17,GWWZ14, GDRS17,BMVPS18,KVPRC20, LM23}. The present numerical results and analysis suggest that the post-pulse dipole instability discovered in recent works~\cite{RDDSV23,RDDVS23,HDDVRS23,HDDRS25,DRDHS26} represents another case where the lack of memory-dependence gives qualitatively wrong behavior. The instantaneous dependence on the density leads to an incorrect non-linearity in the TDKS Hartree-exchange-correlation potential that self-drives small almost pure oscillations in the system after the pulse, until the oscillations grow so as to detune itself through spurious peak-shifting. The CAP plays a critical role: turning it off means that the system is left in a more 'noisy' state, with different frequencies interfering such that the artificial resonant effect does not occur. 

This unphysical behavior vanishes when the very same approximation is used in RR-TDDFT. Exponential growth is not possible within the linear ODE structure of the time-evolution equations in the absence of any field. This example then highlights another advantage of separating out the space and time dependence in TDDFT as RR-TDDFT does: while Ref.~\cite{DBM24} noted this reduces the complexity arising from the inherent weaving of space and time non-localities in the full $v\xc$ functional, we note here that it also removes unphysical behavior arising from the incorrect non-linear dependence of adiabatic functional approximations.

Throughout the dynamics, the closer agreement with CC2 of RR-TDDFT over TDKS using any of the functionals studied here  is consistent with our expectation that even aside from the instability question, the RR-TDDFT framework is more accurate than TDKS with adiabatic functionals because these functionals are evaluated on a domain close to the domain in which they were derived. These functionals miss states of double-excitations, while CC2 contains them although in an approximate  way, so this could be part of the difference in their predictions of the dynamics. While TDKS can access states these states, it does so in an unreliable and inconsistent way~\cite{IL08,RBDM26}. 

Further, we believe that the question of double-excitations is related to divergences that appear in excited-to-excited state couplings from quadratic response TDDFT and CC2~\cite{PRF16,PRF18,DRM23}; the pseudowavefunction approximation we used in the TDDFT couplings avoids these unphysical divergences. However, we found that they did not influence the dynamics of the dipole significantly in the cases studied in this paper: both full quadratic response and even an unrelaxed calculation of the couplings gave similar results for the dipole-dynamics in these cases, suggesting it is likely that within the functionals used, such states were not significantly involved in the dynamics, or at least did not affect the dipole observable much.  

While the results here unequivocally demonstrate that the dipole instability of TDKS vanishes when the same functional is used within RR-TDDFT, the quantitative difference in their predictions while the pulse is on needs further investigation. In particular, a more careful analysis of how a particular CAP used on the molecular orbitals in TDKS should translate over to the CAP on the underlying many-body states in RR-TDDFT is needed and left for future work. 
While highly non-trivial to prove, we do not expect the conclusions regarding the appearance of the instability in adiabatic TDKS and its absence in RR-TDDFT would change in the limit of an exact treatment of the continuum, although we do expect the details of the growth to differ as one moves to more and more realistic continuum treatments, just as they do for the different ways the continuum is treated here and in the previous literature (real-space vs atomic-basis, real-space masking absorber vs energy-decay).

Also left for future work is a deeper analysis beyond that given in Sec.~\ref{sec:why} on the self-driving resonance effect that we believe is at the root of the instability in adiabatic TDKS, including understanding details such as how the onset times depend on the functional as well as on predicting field parameters for which the instability occurs or not. 
Future work would also explore the connection of the dipole-instability of adiabatic TDDFT with the spurious peak-shifting.  These are two distinct symptoms of the same ailment: the instantaneous density-dependence of the adiabatic time-dependent exchange-correlation functional. With $v\xc^A[n;\Psi(0),\Phi(0)](\br,t)= v\xc^{\rm g.s.}[n(\br,t)]$, the spurious peak-shifting effect always arises whenever the density evolves significantly far from a ground state. In contrast, the self-driving effect occurs only under certain conditions, some of which have been  uncovered in Section IIIC. In the initial stages of the dipole growth, we expect the spurious peak-shifting to be minimal. As the oscillations grow and the density changes become larger,  the spurious peak-shifting curbs the growth of the instability, because the system detunes itself. 

Finally, the cases studied here do not model the case where the system begins in an excited cationic state as was a focus in Refs.~\cite{RDDSV23,HDDVRS23,DRDHS26}. What we can learn from comparing RR-TDDFT and TDKS dynamics starting in such a 'drilled hole' state,  especially how the analysis based on the potential energy surface is affected~\cite{DRDHS26}, and whether there  is a difference in the nature of the instability when there is such a population inversion, is left for future study. 

%\begin{table}[h!]
%\centering
%\begin{tabular}{|l|c|c|}
%\hline
%\textbf{Method} & \textbf{electrons left} & %\textbf{electrons ionizaed} \\ \hline
%PBE & 9.4739 & 4.5261 \\ \hline
%AUG/PBE & 11.225 & 2.775 \\ \hline
%Hartree Fock & 9.549 & 4.451 \\ \hline
%%PBE0 & 9.105 & 4.895 \\ \hline
%CC2 & 12.16 & 1.84 \\ \hline
%R2SCAN & 8.44 & 5.56 \\ \hline
%\end{tabular}
%\caption{Method vs Electrons Left and Ionized}
%\label{tab:electrons_data}
%\end{table}
\acknowledgments{
We thank Phuong Mai Dinh, Paul-Gerhard Reinhard, Eric Suraud, and Tchavdar Todorov, for lively and  stimulating discussions about the dipole instability. 
Financial support from the National Science Foundation Award CHE-2154829 (DR) and the Department of
Energy, Office of Basic Energy Sciences, Division of
Chemical Sciences, Geosciences and Biosciences under
Award No. DE‐SC0024496 (NTM) are gratefully acknowledged. The authors acknowledge the Office of Advanced Research Computing (OARC) at Rutgers, The State University of New Jersey for providing access to the Amarel cluster and associated research computing resources that have contributed to the results reported here. }

\bibliography{ref.bib}
\end{document}